\documentclass[preprint,12pt]{elsarticle}

\def\lbldef#1#2{\expandafter\gdef\csname #1\endcsname {#2}}

\def\href#1#2{#2}

\usepackage{color}
\usepackage{graphicx}
\usepackage{amssymb}
\journal{Physics of the Dark Universe}

\begin{document}
	
	\begin{frontmatter}

		\title{The Electroweak Horizon Problem}
		
		\author[1]{Fulvio Melia\footnote{John Woodruff Simpson
				Fellow. E-mail: fmelia@email.arizona.edu}} 
		
		\address[1]{Department of Physics, The Applied Math Program, and Department of Astronomy,
			The University of Arizona, AZ 85721, USA}
		
		\begin{abstract}
Spontaneously broken symmetries in particle physics may have produced
several phase transitions in cosmology, e.g., at the GUT energy scale
($\sim 10^{15}$ GeV), resulting in a quasi-de Sitter inflationary expansion,
solving the background temperature horizon problem. This transition would
have occurred at $t\sim 10^{-36}-10^{-33}$ seconds, leading to a separation
of the strong and electroweak forces. The discovery of the Higgs boson confirms
that the Universe must have undergone another phase transition at the
electroweak (EWPT) scale $159.5\pm1.5$ GeV, about $10^{-11}$ seconds later,
when fermions and the $W^\pm$ and $Z^0$ bosons gained mass, leading to the
separation of the electric and weak forces. But today the vacuum expectation
value ({\it vev}) of the Higgs field appears to be uniform throughout the
visible Universe, a region much larger than causally-connected volumes
at the EWPT. The discovery of the Higgs boson thus creates another serious
horizon problem for $\Lambda$CDM, for which there is currently no established
theoretical resolution. The EWPT was a smooth crossover, however,
so previously disconnected electroweak vacuua might have homogenized as they
gradually came into causal contact. But using the known Higgs potential and
{\it vev}, we estimate that this process would have taken longer than the age
of the Universe, so it probably could not have mitigated the emergence of
different standard model parameters across the sky. The EWPT horizon problem
thus argues against the expansion history of the early Universe predicted by
standard cosmology.
\end{abstract}
		
	\end{frontmatter}
	
\section{Introduction}
The impact of particle-physics-induced phase transitions on cosmology
was recognized over four decades ago, e.g., via the proposal in the
early 1980's of solving the cosmic microwave (CMB) temperature horizon 
problem using the quasi-de Sitter expansion produced when the strong 
and electroweak forces separated at the grand unified (GUT) scale 
$\sim 10^{15}$ GeV \cite{Starobinskii:1979,Kazanas:1980,Guth:1981,Linde:1982}. 
But this could not have been the only spontaneously broken symmetry impacting
the expansion of the Universe. Aside from this well-studied case, which was 
actually originally motivated by missing magnetic monopoles, there should have 
been at least one more associated with the separation of the electric and weak
forces \cite{Glashow:1961,Weinberg:1967,Salam:1968}.

We know this with confidence following the discovery of the Higgs particle 
\cite{Aad:2012}, whose existence confirms the widely held belief that inertial
mass in the standard model is at least partially due to the Higgs mechanism
\cite{Englert:1964,Higgs:1964}. Thus, a second well-motivated transition (the 
electroweak phase transition, EWPT) must have occurred at a critical temperature 
of $159.5\pm 1.5$ GeV.  But as we shall see, this creates a problem because in 
$\Lambda$CDM this temperature would have been reached $\sim 10^{-11}$ seconds 
after the Big Bang, when causally-connected regions were still too small to 
eventually fill the Universe we see today. On the other hand, the EWPT would 
have occurred well past the first (inflationary) transition at $t\sim 10^{-36}
-10^{-33}$ seconds, so its impact could not have been mitigated by the 
quasi-de Sitter expansion completed earlier. 

In the standard model of particle physics, the EWPT is a `crossover' (always close 
to equilibrium), rather than first order (marked by a discontinuity). In principle, 
the latter could have produced the baryon asymmetry observed in matter left 
over after particle annihilations ended as the Universe cooled (see, e.g., 
ref.~\cite{Cline:2018} for a recent summary). The promise of such a scenario
motivates possible extensions to the standard model in order to circumvent the 
implied limitations of a crossover. These include the introduction of additional 
Higgs fields \cite{Fileviez:2009}, that could also generate gravitational waves 
(see, e.g., \cite{Weir:2018}) detectable with LISA \cite{Caprini:2016,Audley:2017}. 
The search for hints of a Higgs self-interaction consistent with these models 
thus continues with the High-Luminosity Large Hadron Collider (LHC) (see, e.g., 
\cite{Noble:2008,Dolan:2012,Barr:2015}), and will be featured in future particle
accelerator experiments. Currently, however, a primary focus on Higgs tends to 
be the aforementioned generation of fermionic, $W^\pm$ and $Z^0$ mass, believed 
to have separated the electric and weak forces---a crucial event in the history 
of the Universe.

Of course, if the standard model is correct, a third phase transition should
have occurred at $\sim 100$ MeV, some $10^{-6}$ seconds after the EWPT. In 
quantum chromodynamics (QCD), such an event would have ensued following the 
condensation of free quarks in a quark-gluon plasma into the confined 
states representing baryons and mesons as the Universe continued to expand 
and cool.

Our principal concern in this paper is the inevitable horizon problem
created by the Higgs mechanism in the context of $\Lambda$CDM, analogously
to the earlier horizon problem associated with the CMB temperature. This
time, however, the pertinent physical quantity is the vacuum expectation
value ({\it vev}) of the Higgs field, which is apparently uniform throughout
the Universe, even on scales greatly exceeding regions that could not have
been causally-connected at the time of the EWPT. As we shall see, while
inflation might have mitigated the temperature horizon problem, its
implied de Sitter expansion would have occurred well {\it before} the
EWPT, and would thus have been largely irrelevant to the emergence of
the Higgs {\it vev}.

Currently, there is no established solution to this problem, which has slowly 
gained in prominence over the past half century, culminating with the recent
experimental confirmation of the Higgs mechanism. In one of its earliest guises, 
the EWPT was thought to create sub-horizon features, manifested as observable 
anisotropies in the CMB. For example, Zel’dovic, Kobzarev \& Okun 
\cite{Zeldovic:1975} and Kibble \cite{Kibble:1976} assessed the possibility 
that `domain walls' might have been created from such phase 
transitions in the early Universe. Some measurable features in the CMB could
in principle be associated with these `topological defects' 
\cite{Vilenkin:1994,Lazanu:2015,Sousa:2015}. Five decades later, however, 
we have a much more detailed understanding of the CMB temperature fluctuations,
and we have seen no evidence of domain walls created by the EWPT 
\cite{Hinshaw:1996,Bennett:2003,Planck:2020}.

Below, we shall first provide a brief background on the Higgs mechanism, followed
by a quantitative demonstration of the EWPT horizon problem.  We will conclude with 
a discussion of some attempts made thus far to address this quandry which, however, 
is now much more serious and better established than ever before following the 
discovery of the Higgs boson. 

\section{Background}
With the recent discovery of the Higgs boson, the Higgs mechanism for 
generating the inertia of fermions, and the $W^\pm$ and $Z^0$ bosons is 
now widely accepted \cite{Englert:1964,Higgs:1964}. The two principal 
issues in this process are (i) when did the Higgs field acquire a 
non-zero {\it vev}, commonly referred to as `turning on the Higgs field'? 
and (ii) how strong is the Higgs coupling to the various elementary 
particles?  The electroweak symmetry is unbroken at asymptotically 
high temperatures because all of the `messenger' particles ($W^\pm$, $Z^0$ 
and photons) mediating the electroweak force transfer the same momentum 
per unit energy from one fermion to the next. From the relativistic 
expression for energy, $E^2=m^2c^4+p^2c^2$, one can see that the value 
of $p/E$ becomes independent of $m$ in the regime where $p\gg mc$.

This symmetry is spontaneously broken, however, and the electric and
weak forces separate, when $p/E$ becomes dependent on the particle
type due to the emergence of a non-zero mass. In the absence of a Higgs
mechanism, this would happen gradually as the Universe cooled to a 
temperature $T\sim m_\alpha c^2/k_{\rm B}$, where $m_\alpha$ is the 
$W^\pm$ or $Z^0$ mass and $k_{\rm B}$ is the Boltzmann constant. If
the particle rest-mass energy is directly due to the Higgs
coupling, however, the transition would have happened when the Higgs 
field acquired a non-zero vev. The viability of the Higgs mechanism 
makes the spontaneous symmetry breaking cleaner and more precisely 
localized in temperature and time. And as noted earlier, we now know
that the EWPT must have occurred at $k_{\rm B}T=159.5\pm1.5$ GeV, 
the temperature to which the $\Lambda$CDM universe would have cooled 
by $t\sim 10^{-11}$ seconds. 

Clearly, whether the EWPT created a horizon problem or not thus
depends on the value at which the Higgs field finally settled. But 
we have no reason to believe that the {\it vev} is specified uniquely.
In principle, it could have been anything. What we can 
say for certain, however, is that whatever conditions established the 
{\it vev}, it would have been created uniformly only throughout a 
causally-connected region of spacetime. There is no known initial 
constraint that could otherwise have forced the Higgs field to emerge 
with the same value even at distances exceeding our causal horizon.
Note in particular, that the {\it vev} is in fact associated with 
an operator which, in quantum mechanics would be independent of the 
observer only if `hidden variables' were to establish its magnitude
in terms of preset physical conditions. But many modern tests of Bell's
theorem have compellingly shown that hidden variables almost certainly
do not exist \cite{Hensen:2015,Giustina:2015,Shalm:2015}.

An interesting approach to this question is based on an anthropic
constraint for the existence of atoms \cite{Donoghue:2010}, which 
allows one to estimate the likelihood function for the Higgs {\it vev}.
The fermionic masses are proportional to the Higgs {\it vev}, the
argument goes. Thus, since nuclei and atoms could only exist if the 
light-quark and electron masses were close to their actual measured 
values \cite{Agrawal:1998,Hogan:2000,Damour:2008}, the anthropically 
permitted bounds constitute constraints on the {\it vev} distribution 
producing observers, subject to  the other parameters in the 
standard cosmological model.

Without such anthropic considerations, there is actually quite a large 
domain of possible {\it vev}'s, due to an unknown property of the 
fundamental theory, extending at least up to the GUT scale, orders of 
magnitude greater than the current EW scale. Indeed, the present 
disparity between the GUT and EW scales constitutes a so-called 
`hierarchy problem.' One may thus explore how the {\it vev} 
distribution is shaped by variations in the cosmology, but 
always mindful of the requirement that nuclei and atoms must 
appear.

As it turns out, the range of {\it vev}'s permitted by this anthropic 
analysis is far smaller than the EW to GUT gap, but it is nevertheless
not minute. Rather than the Higgs field being constrained solely to its
measured value, referred to as $v_0$, the {\it vev} distribution 
actually has a median value of $2.25v_0$, with a $2\sigma$ range 
extending from $0.10v_0$ to $11.7v_0$. Thus, since fermionic masses 
are proportional to the {\it vev}, the nuclear and atomic properties 
we observe in the real Universe could have varied by over two orders 
of magnitude, from one causally-connected region to another. As noted 
earlier, however, we have never seen such variations over the past 
half-century of observations, culminating with the latest, most 
precise measurements carried out by {\it Planck} \cite{Planck:2020}.

\section{The Electroweak Horizon Problem}
The most straightforward way to see why the EWPT creates a serious 
horizon problem for standard cosmology is the following. We write the 
Hubble parameter for flat $\Lambda$CDM in the form
\begin{equation}
H(a)=H_0\sqrt{\Omega_{\rm m}\,a^{-3}+\Omega_{\rm r}\,a^{-4}+\Omega_\Lambda}\;,\label{eq:Ha}
\end{equation}
where $a(t)$ is the expansion factor in the Friedmann-Lema\^itre-Robertson-Walker (FLRW)
metric, and the Hubble constant ($H_0=67.8$ km s$^{-1}$ Mpc$^{-1}$) and the scaled
densities for matter ($\Omega_{\rm m}=0.308$), radiation ($\Omega_{\rm r}=
5.37\times 10^{-5}$) and dark energy ($\Omega_\Lambda=1-\Omega_{\rm m}-\Omega_{\rm r}$),
take on their {\it Planck} values \cite{Planck:2020}. For this model, the redshift 
and age at decoupling were $z_{\rm dec}=1089.9$ and $t_{\rm dec}=377,700$ years, 
respectively. Thus,  
\begin{equation}
a(t_{\rm dec})=(1+z_{\rm dec})^{-1}\approx 9.17\times 10^{-4}\;.
\end{equation}
The corresponding Hubble parameter (from Eq.~\ref{eq:Ha}) was thus
\begin{equation}
H(t_{\rm dec})\approx 4.78\times 10^{-14}\;\, {\rm s}^{-1}\;.
\end{equation}
The gravitational radius (coincident with the size of the Hubble sphere) at that 
time may thus be calculated as \cite{Melia:2018} 
\begin{equation} 
R_{\rm h}(t_{\rm dec})\equiv {c\over H(t_{\rm dec})}\approx 0.20\;\,{\rm Mpc}\;.
\end{equation}

This allows us to infer the expansion factor, $a(t)$, and Hubble parameter,
$H(t)$, at any time $t$ prior to decoupling from the definition 
\begin{equation}
t_{\rm dec}-t=\int_{a(t)}^{a_{\rm dec}}{da\over a\,H(a)}\;.
\end{equation}
Setting $t=t_{\rm ew}=10^{-11}$ seconds, thus allows us to solve for $a(t_{\rm ew})$ at 
the EWPT, yielding 
\begin{equation}
a(t_{\rm ew})\sim 10^{-15}\;,
\end{equation}
with a corresponding gravitational radius 
\begin{equation}
R_{\rm h}(t_{\rm ew})\sim 1.3\;{\rm cm}\;.\label{eq:Rhew}
\end{equation}

From previous studies \cite{Melia:2018,Bikwa:2012,Melia:2012,Melia:2013},
we know how to determine the proper size of a causally-connected region in
terms of $R_{\rm h}$. Unlike the situation in a static spacetime, like 
Schwarzschild, $R_{\rm h}$ in the cosmological context is not an event 
horizon. In general relativity this quantity represents a so-called
`apparent' horizon that separates null geodesics approaching the observer
from those that are receding \cite{Melia:2018}. $R_{\rm h}$ may turn
into an event horizon in our distant future, depending on the cosmic
equation of state, but for now we recognize that our gravitational (or
apparent) horizon is changing with time, so the congruence of null geodesics
reaching us at any given time also changes as the Universe expands. Consequently, 
regions that were beyond our causal horizon in the past, can enter into
our current causally-connected portion of the Universe.

In simplest terms, the key physical determinant of whether or not a 
distant source is causally connected to us is whether or not a light
signal it emitted in the past has reached us by today. And in this context,
previous studies \cite{Melia:2018,Bikwa:2012,Melia:2012,Melia:2013} have 
shown that an observer receiving a light signal at time $t_{\rm obs}>t_{\rm max}$
infers a maximum photon excursion $R_{\gamma}(t_{\rm max})\lesssim 
R_{\rm h}(t_{\rm obs})/2$ away from them. The time $t_{\rm max}$ defines 
the point on the observer's past lightcone at which $R_{\gamma}$ is maximized.  

This behavior of null geodesics, in terms of the proper distance $R_\gamma(t)$ 
in FLRW, is not difficult to understand \cite{Melia:2013}. de Sitter space 
is the only well known FLRW model with a time-independent metric and 
no initial singularity. All other models, including $\Lambda$CDM, began 
their expansion at a specific time (i.e., the Big Bang), and therefore
could not have had pre-existing, detectable sources lying away from the observer's
location prior to the Big Bang. Except in de Sitter, all of the photons detected 
by the observer at time $t_{\rm obs}$ from the most distant, obervable locations 
were emitted {\it after} their sources had reached the edge of visibility---at a 
proper distance of roughly $R_{\rm h}(t_{\rm obs})/2$.  This defines the proper 
size of the {\it visible} Universe at any given time $t_{\rm obs}$, (see 
ref.~\cite{Melia:2018} for a more detailed description).

Some authors have claimed that today (at time $t_0$) we can see sources beyond 
$R_{\rm h}(t_0)$ (see, e.g., ref.~\cite{Davis:2001}). But these papers are 
confusing the location of the sources today with where they were when they 
emitted the light we are just now receiving. We certainly cannot expect that 
our causally-connected region is defined by light signals we shall receive in
our future. Causal contact must be established within the proper size 
$R_\gamma(t_{\rm max})$ by light signals that have actually already been 
exchanged between the emitter and the observer.

Shifting forward to the present, we estimate that
\begin{equation}
R_{\rm ew}(t_0)\equiv \left[{a(t_0)\over a(t_{\rm ew})}\right]R_{\rm h}(t_{\rm ew})\approx 
10^{-3}\; {\rm lyr}\;.\label{eq:Rew}
\end{equation}
Thus, to within a factor $\sim 2$ \cite{Melia:2013}, this is the size today of the 
region that was causally-connected at the EWPT. It therefore represents the 
largest region within which the Higgs {\it vev} ought to be uniform. The horizon
problem arises because the gravitational radius of the Universe is now 
$R_{\rm h}(t_0)\approx 4,424$ Mpc---many orders of magnitude larger than this. 
If the expansion history in $\Lambda$CDM were correct, the fermionic and atomic 
properties we observe around us should thus be varying spatially across the Universe, 
as one would expect from the aforementioned topological defects discussed by Zel’dovic, 
Kobzarev \& Okun \cite{Zeldovic:1975} and Kibble \cite{Kibble:1976} half a century ago.

\section{Discussion}
To place the radius $R_{\rm ew}(t_0)$ in context, compare $R_{\rm h}(t_{\rm ew})$ in 
Equation~(\ref{eq:Rhew}) with the corresponding gravitational radius at the start of 
inflation, presumably at $t_{\rm inf} \sim 10^{-36}$ seconds after the Big Bang. The 
Universe would have been dominated by radiation for $t\lesssim t_{\rm ew}$, so we can 
put $a(t)\propto t^{1/2}$ and $H(t)\propto t^{-3/4}$ during this epoch. This yields 
$R_{\rm h}(t_{\rm inf})\sim 1.6\times 10^{-25}$ cm. Of course, with the hypothesized 
subsequent $60$ e-folds of expansion before the Universe settled back into its hot Big 
Bang dynamics, this radius would have increased almost instantaneously to $\sim 16$ cm, 
already larger than the electroweak horizon radius $R_{\rm h}(t_{\rm ew}) \sim 1.3$ cm 
about $10^{-11}$ seconds later. One can easily estimate from this that the inflated 
causally-connected volume at the end of inflation would thus have been larger than
the whole Universe we see within our gravitational radius today, i.e., 
$R_{\rm inf}(t_0)\equiv [a(t_0)/a(t_{\rm inf})]R_{\rm h}(t_{\rm inf})>R_{\rm h}(t_0)$.

\subsection{A Second Inflationary Expansion}
The obvious question is therefore whether an analogous spurt of inflated expansion could
have happened a second time to rescue the glaring inconsistency implied by Equation~(\ref{eq:Rew}). 
A milder, delayed inflationary phase has in fact been proposed on several occasions, for 
various reasons not necessarily having to do with the EWPT itself
\cite{Randall:1995,Lyth:1996,Randall:1996,German:2001,Boeckel:2010,Davoudiasl:2016}. 
For example, models of supersymmetry breaking contain many possible scalar and fermionic 
fields, loosely referred to as `moduli,' characterized by masses of order the weak scale 
and gravitational-strength couplings to the visible sector. Their corresponding quanta
could have posed a serious problem to cosmology, however, if produced in the early Universe.
They would have behaved like nonrelativistic matter, decaying very slowly, and dominating 
the cosmic energy density past the end of nucleosynthesis. Their decay products could have
destroyed $^4$He and D nuclei by photodissociation, thus ruining any hope of a working BBN 
model within standard cosmology \cite{Coughlan:1983,Ellis:1986,deCarlos:1993,Banks:1994}. 
An example of such a relic is the supersymmetric partner of the graviton, called the
spin-3/2 gravitino. 

This so-called cosmological moduli problem may be solved, however, with a period of weak-scale
inflation \cite{Randall:1995,Lyth:1996,German:2001}, which could have diluted the density of 
moduli below their destructive level. In this proposal, the hot Big Bang may not have persisted 
uninterrupted through the electroweak scale if one or more of these hypothesized scalar fields 
had a sufficiently large vacuum expectation value to temporarily dominate the energy density 
with an almost flat potential. In other words, this inflationary process would have arisen 
`naturally' from the same assumptions that lead to the cosmological problem in the first place. 

But unlike the GUT inflationary phase invoked to solve the CMB temperature horizon
problem, this late-time, weak-scale inflation would have produced a mere $\sim 10$
e-folds of inflation, growing the size of a typical, causally-connected electroweak 
region to $\sim 2.2\times 10^4 R_{\rm ew}(t_0)\approx 22$ lyr---still many orders of
magnitude below the gravitational radius of the Universe today.

A similar mechanism for a late-time inflationary phase has also been proposed at
the QCD phase transition \cite{Boeckel:2010}. In this scenario, the phase transition
would have signaled the transformation in the early Universe from a quark-gluon plasma to
a hadron gas at a critical temperature $T_{\rm QCD}\approx 150-200$ MeV, corresponding
to a time $\sim 10^{-6}-10^{-4}$ seconds in the standard model. But this scenario, it
turns out, is very similar to the late-time weak-scale inflation, with a length
of only $\sim 10$ e-foldings. Thus, while it may have mitigated other cosmological 
problems, it cannot even come close to solving the electroweak horizon problem.

Finally, a more generic form of late-time inflation has been proposed to account
for the fact that dark matter candidates today, such as the weakly interacting massive
particles (WIMPs) or the QCD axion \cite{Weinberg:1978,Wilczek:1978}, appear to have 
a much lower density than one might expect from their overproduction in the early
Universe \cite{Davoudiasl:2016}. To `fix' this particular problem, an imprecisely
defined scenario referred to as `inflatable dark matter' has been proposed, in which 
a brief period of late-time inflation could have occurred with an energy scale from 
several MeV to hundreds of GeV. 

As noted earlier, however, such an inflationary event would require a source of
vacuum energy exceeding the radiation energy density in the early Universe, at least
briefly. We know of at least some possible candidates, including those associated
with the QCD and electroweak phase transitions. But as we have seen, these cases
could not have provided a sufficient number of e-foldings to mitigate our problem.
And it now appears that these fields transitioned to their broken phase before 
they could dominate the energy density, so they probably could not have provided 
the late-time inflation anyway \cite{Davoudiasl:2016}.  Thus, to conceive of a second 
inflationary event that might also have solved the electroweak horizon problem, one 
would need to postulate a period of inflation triggered by the potential energy 
density of fields beyond the Standard Model.

It remains to be seen whether such extensions can eventually be made to work, 
however.  Unlike `standard' inflation that may have possibly occurred very early 
in the Universe's history, late-time inflation would have been uncomfortably close
to other physically important events---such as BBN---requiring a fine tuning of
conditions to fix the electroweak horizon problem, while not breaking other
aspects of the cosmic expansion that seem to be currently viable. We note,
in this regard, that the new inflationary field would need to have had a 
potential dominating the energy density for more than 30 e-foldings 
in order to solve the electroweak horizon problem. But such a dramatic 
expansion just prior to (or during) baryogenesis and BBN would probably have 
broken the concordance model, not to mention overly diluting any dark matter 
candidates to densities below observationally relevant levels, defeating the
purpose for which such fields were proposed in the first place. 

\subsection{Homogenization of causally disconnected electroweak vacuua}
If not another inflationary spurt, one might contemplate the mitigation
of the EWPT horizon problem via the homogenization of different electroweak 
vacuua as they gradually came into causal contact. Since the EWPT is now 
known to have been a crossover, one would not expect topologically stable 
domain walls to have formed, allowing different vacuua to be mutually 
accessible. Gradients in the {\it vev} could thus have affected the
Higgs-field dynamics, possibly establishing a uniform {\it vev} across
the observable Universe. 

We shall adopt a highly simplified approach to estimate the timescale
over which this process could have acted, starting with the equation of 
motion for the Higgs field in the FLRW metric (see, e.g., 
ref.~\cite{Englert:1964,Higgs:1964,Weinberg:1978,Melia:2020}):
\begin{equation}\ddot{\phi}+3H(t)\dot{\phi}+V^\prime(\phi)=0\;,\label{eq:dyn}
\end{equation}
where $H(t)$ is the Hubble parameter and the Higgs potential may be written
\begin{equation}
V(\phi)={\lambda\over 4}\left(\phi^2-v^2\right)^2\;,\label{eq:pot}
\end{equation}
in terms of the self-interaction coupling of the Higgs boson, $\lambda$, and
the {\it vev}, $v$, which today has a measured value of $246$ GeV. The Higgs 
boson mass itself is given by $m\equiv \sqrt{2\lambda}v$.

We focus on small background Higgs field values, near the minimum of the potential,
so that we may neglect the quadratic term. In addition, we ignore for simplicity
the contribution from local spatial gradients in $\phi$ itself, though these would
clearly contribute to the global evolution in the {\it vev} (see Eq.~\ref{eq:vdyn}
below). It is not difficult to see that, in this case, we may approximate the 
potential with the simpler expression
\begin{equation}
V(\phi)\approx {1\over 2}m^2(\phi-v)^2\;,\label{eq:potsim}
\end{equation}
which then gives
\begin{equation}
V^\prime(\phi)\approx m^2(\phi-v)\;.\label{eq:vprim}
\end{equation} 
We suppose that gradients in the {\it vev} (within regions where previously
disconnected vacuua come into causal contact) would manifest themselves via
a time-dependent $v$, but always follow the field near the minimum of $V(\phi)$, 
as described above. In that case, $\dot{\phi}\sim \dot{v}$, and Equation~(\ref{eq:dyn})
simplifies to
\begin{equation}
\ddot{v}+3H\dot{v}+m^2k v\approx 0\;,\label{eq:vdyn}
\end{equation}
where we take $k$ to be a constant of order $\lesssim 1$.

In the standard model, $H(t)$ averaged over a Hubble time is remarkably
close to $1/t$, an odd coincidence that no doubt points to some fundamental
physics \cite{Melia:2020}, though we do not need to explore that here. It
does, however, allow us to simplify Equation~(\ref{eq:vdyn}) even further,
so that
\begin{equation}
\ddot{v}+{3\over t}\dot{v}+m^2kv\approx 0\;,\label{eq:vdyn2}
\end{equation}
which has the trivial solution $v(t)=v_{\rm init}/t$, with $v_{\rm init}
\equiv 1/\sqrt{2k\lambda}$.

Suppose then that adjacent electroweak vacuua had a {\it vev} mismatch by
a factor $n\lesssim 10$, consistent with our expectation from the anthropic
principle discussed in \S~2 above. At what time, $t_{\rm init}$, would the 
homogenization process need to have started in order for us to see a uniform 
$v=246$ GeV today?  According to the simple solution to Equation~(\ref{eq:vdyn2}),
\begin{equation}
t_{\rm init}\sim {1\over \sqrt{k}nm}\;.\label{eq:tinit}
\end{equation}
Both $k$ and $n$ are of order 1, so $t_{\rm init}\sim 0.008$ GeV$^{-1}$, which
translates to a time $\sim 5\times 10^{-23}$ seconds. But the EWPT must have
occurred much later, at $t\sim 10^{-11}$ seconds, so the Higgs {\it vev} 
probably could not have become homogenized across the observable Universe today.

\vskip 0.1in
\section{Conclusion}
No reliable evidence has ever been found of a breakdown in the physical
properties of atomic and nuclear matter on cosmic scales (see, e.g., 
ref.~\cite{Planck:2020}). Moreover, all of the structure we have seen 
throughout the visible Universe appears to be made of matter, not antimatter. 
So if the baryon asymmetry is due to the EWPT, as some have suggested
via the introduction of additional Higgs fields, its uniformity affirms
the measurement of a uniform Higgs {\it vev} throughout our causally-connected
spacetime. 

But correspondingly, no viable solution to the electroweak horizon problem
has been proposed either. Given the recent discovery of the Higgs boson,
and its implied confirmation of the Higgs mechanism for generating
inertia in the Standard Model, there is no question now that the
culpability for any conflict between the EWPT and cosmology must be
placed squarely upon the latter. It is becoming increasingly clear that
the most likely resolution of this problem is to avoid it altogether,
calling for a major overhaul of the physical basis for predicting
the expansion history in $\Lambda$CDM \cite{Melia:2020}.  

\section*{Acknowledgments} 
I am very grateful to the anonymous referee for an excellent suggestion
to improve the presentation in this manuscript.

\end{document}